\documentclass[twocolumn,prb,aps,amssymb]{revtex4}
\usepackage{graphicx}
\usepackage{dcolumn}
\usepackage{bm}

\begin{document}

\title{
Effective interactions in colloid - semipermeable membrane systems
}
\author{Pawe{\l} Bryk}
\affiliation{Department for the Modeling of Physico-Chemical Processes,
Maria Curie-Sk{\l}odowska University, 20-031 Lublin, Poland}
\email{pawel@paco.umcs.lublin.pl}
\date{\today}
\begin{abstract}
We investigate effective interactions between a colloidal particle,
immersed in a binary mixture of smaller spheres,
and a semipermeable membrane. The colloid is modeled as a big hard sphere
and the membrane is represented as an infinitely thin surface which
is fully permeable to one of the smaller spheres and impermeable to the other
one. Within the framework of the density functional theory we evaluate the
depletion potentials, and we consider two different approximate
theories - the simple Asakura-Oosawa
approximation and the accurate White-Bear version of the fundamental measure
theory. The effective potentials are compared with the corresponding
potentials for a hard, nonpermeable wall. Using statistical-mechanical sum
rules we argue that the contact value of
the depletion potential between a colloid and a semipermeable membrane
is smaller in magnitude than the potential between a colloid
and a hard wall.
Explicit calculations confirm that the colloid-semipermeable membrane
effective interactions are generally weaker than these near a hard nonpermeable wall.
This effect is more pronounced for smaller osmotic pressures.
The depletion potential for a colloidal particle inside a semipermeable
vesicle is stronger than the potential for the colloidal particle located
outside of a vesicle. The asymptotic decay of the depletion potential for
the semipermeable membrane is similar to that for the nonpermeable wall and
reflects the asymptotics of the total correlation function of the
corresponding binary mixture of smaller spheres. Our results demonstrate
that the ability of the membrane to change its shape constitutes
an important factor in determining the effective interactions between the
semipermeable membrane and the colloidal macroparticle.
\end{abstract}

\maketitle
\section{Introduction}
Depletion interactions play an important role in determining the
stability of colloidal systems \cite{Likos01,Poon02}.
They arise in asymmetric mixtures
\cite{Tuinier03} due to a tendency to maximize volume available to centers
of smaller species.
In suitably prepared colloidal suspensions, such that only bare  hard-core
interactions are left, these effective interactions are attractive at small distances
and can exhibit a potential barrier and an oscillatory tail for large
separations \cite{Dickman97,Biben96}.
Depletion forces also arise if a big colloidal particle  
suspended in a sea of small particles approaches a fixed object such as
a planar wall.

The first theoretical approach to calculate the effective interactions was
performed by Asakura and Oosawa \cite{Asakura54}, and later, by Vrij
\cite{Vrij76}.
The resulting effective potential is attractive at small distances and
proportional to the packing fraction of the depleting agent.
More recent approaches include various integral equation
\cite{Attard90,Attard92,Amokrane98,Kinoshita02} and density functional
theories \cite{Goetzelmann98,Goetzelmann99,Roth00}. 
Several experimental techniques have been developed to investigate depletion
interactions. Total internal reflection microscopy has been
used to investigate depletion potentials between a wall and a sphere immersed
in a sea of spherical polymers \cite{Bechinger99} and rod-like particles
\cite{Helden03,Roth03}.
Video-microscopy has been used to measure the depletion potential  between two big
particles \cite{Crocker99}, the potential of a big colloid near an edge of a terrace
\cite{Dinsmore96} and the potential of a big colloid sphere inside a vesicle \cite{Dinsmore98}.
In the latter experiment it has been observed
that the big particle is attracted to regions of high local curvature.
This attraction may in turn induce transformation of the shape of a membrane.
Thus the depletion interactions between a colloidal particle and a membrane
may be helpful in a better understanding of many biological
processes.

Theoretical attempts to describe the colloid-vesicle experiment 
were initiated by Roth {\it et al}. \cite{Roth99}
who used density functional theory (DFT)
to study depletion forces between a big hard sphere, immersed in a
sea of smaller spheres, and a hard, nonpermeable spherical wall.
These authors also presented a simple framework for tackling cavities
of a general shape. A more refined approach was presented very recently
\cite{Koenig04,Koenig05}.
Bickel investigated depletion interactions near soft, fluctuating surfaces and
showed a mechanism for an encapsulation of the colloidal macroparticle
\cite{Bickel03}.
In all these studies the membrane itself is assumed to be nonpermeable.
However, an inherent attribute of a semipermeable membrane system is
the presence of a fluid on both membrane sides. 
The difference in composition of the fluid on both sides, which appears
due to the selective permeability of a membrane, gives rise to the osmotic pressure.

Several microscopic models of semipermeable membrane systems
have been proposed in the literature \cite{Murad00}. The simplest model is 
the so-called ideal membrane - an infinitely thin surface which is fully
permeable to some of the species and nonpermeable to others.
The fluid structure near such a membrane can be ascertained by means of
first \cite{Zhou88a,Zhou88b} and second order \cite{Bryk97}
integral equation theories or DFT \cite{Bryk98a}. 
A somewhat more sophisticated model accounts
for finite thickness by treating the membrane as a
potential barrier of a finite height and width \cite{Marsh95,Margaritis97,Bryk99}.
Finally, in some studies the discrete structure of a membrane was also
taken into account by considering membranes built of particles arranged
either randomly \cite{Bryk98b} or in a regular manner \cite{Murad97,Murad98,Borowko98,Jia05}.

In the present paper we investigate the effective interactions between
a semipermeable membrane and a big colloidal macroparticle. We use the reliable
DFT approach that was used previously to study depletion potentials near impenetrable
surfaces and inside hard-sphere cavities. This enables us to make a direct
comparison and elucidate the effect of the semipermeability on the effective interactions.
The paper is arranged as follows: in Section \ref{sec:2} we consider sum rules
for semipermeable membranes. Section \ref{sec:3} outlines the theories applied
in the present paper. Section \ref{sec:4} presents results and we conclude in
Section \ref{sec:5}.

\section{Sum rules for semipermeable membranes}
\label{sec:2}
\begin{figure}[t]
\includegraphics[clip,width=8cm]{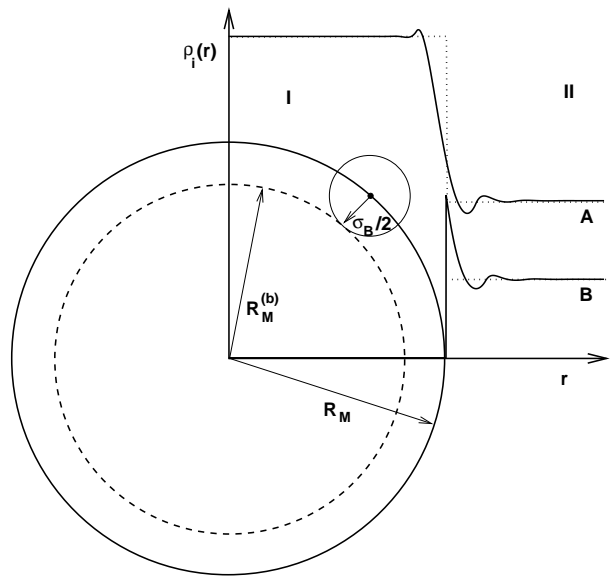}
\caption{\label{fig:1}
Schematic plot of a semipermeable vesicle system.
}
\end{figure}

Consider two component mixture of species $A$ and $B$ and diameters $\sigma_A$
and $\sigma_B$, respectively. The membrane is fully permeable to the component
$A$ but is nonpermeable to the component $B$. In the course of the derivation
we consider the ideal membrane of the spherical shape - a simple model for a
semipermeable vesicle. The sum rule for semipermeable planar membrane is
then obtained by taking the limit $R_M\to\infty$. The membrane is modeled by
imposing an external potential of the form
\begin{equation}\label{eq:1}
V^{ext}_A(r)\equiv 0\; ; \;\;V^{ext}_{B}(r)=\left\{
\begin{array}{ll}
0, & r>R_{M} \\
\infty, &  r<R_{M} \;.
\end{array}
\right.
\end{equation}
In Eq.~(\ref{eq:1}) $R_M=R_M^{(b)}+\sigma_B/2$, and  $R_M^{(b)}$ denotes the
actual membrane radius (the dashed circle in Fig.~\ref{fig:1}). 
$R_M$ stands for the radius at which the profile of the
nonpermeable species makes a jump (the solid circle in Fig.~\ref{fig:1}).
The membrane separates the 
two parts of the system: part II with a two-component fluid 
and part I, available only to the permeable component.
The bulk densities in part II,
$\rho_{A}^{(II)} $
and $\rho_{B}^{(II)}$,
are the input to the calculations. The system is in the termodynamical but
not mechanical equilibrium. The pressure $p$ is different in the two
parts of the system and we define the osmotic pressure as
\begin{equation}\label{eq:2}
\Pi=p^{(II)}(\rho_A^{(II)},\rho_B^{(II)})-p^{(I)}(\rho_A^{(I)}).
\end{equation}
The bulk density in the one-component part of the system, $\rho_A^{(I)}$,
is obtained by imposing the equality of the chemical potentials
of the permeable component in both parts of the system
\begin{equation}\label{eq:3}
\mu_A^{(II)}(\rho_A^{(II)},\rho_B^{(II)})=\mu_A^{(I)}(\rho_A^{(I)}).
\end{equation}
It is clear that the osmotic pressure defined by Eqs.~(\ref{eq:2}) 
and (\ref{eq:3}) depends only on the equation of state of the system
and is independent of the shape of the membrane.

We decompose the grand canonical potential of the system $\Omega$ 
into a reference, bulk part $\Omega^{bulk}=-p^{(II)}V^{(II)}-p^{(I)}V^{(I)}$,
where $V^{(I)}$ and $V^{(II)}$ are the volumes of parts I and II,
respectively, and to an excess part $\Omega^{ex}$:
$\Omega=\Omega^{bulk}+\Omega^{ex}$. We choose $R_M$ and the corresponding
spherical surface of the area $4\pi R_M^2$ as the dividing interface.
This implies that the volumes of both parts are: $V^{(I)}=4/3\pi R_M^3$,
and $V^{(II)}=4/3\pi(\overline{R}^3-R_M^3)$, whereas the densities in
the reference system are
\begin{eqnarray}\label{eq:4}
\rho_A^{bulk}&=&\rho_A^{(I)}\Theta(R_M-r)+\rho_A^{(II)}\Theta(r-R_M)\nonumber\\
\rho_B^{bulk}&=&\rho_B^{(II)}\Theta(r-R_M)\;,
\end{eqnarray}
where $\theta$ is the Heaviside unit-step function
(cf. dotted lines in Fig.~\ref{fig:1}). $\overline{R}$ is a macroscopically
large radius considered in the thermodynamical limit.
Denoting the excess grand potential per unit are as the surface tension
$\gamma$, we have
\begin{equation}\label{eq:5}
\Omega=\gamma(R_M)4\pi R_M^2 -p^{(II)}V^{(II)}(R_M)-p^{(I)}V^{(I)}(R_M)\;.
\end{equation}

We derive the sum rules for semipermeable membranes by expressing
the grand potential as a functional of the local densities
\begin{equation}\label{eq:6}
\Omega[\rho_A,\rho_B]= F[\rho_A,\rho_B]+\sum_{i=A,B}\int d^3 r\rho_i({\bf r})
(V^{ext}_{i}({\bf r})-\mu_i)\;,
\end{equation}
where $F$ is the Helmholtz free energy functional that 
will be defined later. The only explicit dependence of $\Omega$ on $R_M$
is contained in the second term in the right-hand side of Eq.~(\ref{eq:6}). 
At thermodynamical equilibrium the grand potential satisfies the 
Euler-Lagrange equation
\begin{equation}\label{eq:7}
\frac{\delta \Omega [\rho_A,\rho_B]}{\delta \rho_A({\bf r})}=
\frac{\delta \Omega [\rho_A,\rho_B]}{\delta \rho_B({\bf r})} = 0 \;.
\end{equation}

The sum rules are obtained by taking the derivative of the grand potential
Eq.~(\ref{eq:6}) with respect to $R_{M}$, at constant chemical potentials
$\mu_i$ and temperature $T$
\begin{eqnarray}\label{eq:8}
\left(\frac{\partial \Omega}{\partial R_M}\right)_{\mu_i,T}=
\sum_{i=A,B}\int d^3 r 
\frac{\delta \Omega[\rho_A,\rho_B]}{\delta \rho_i({\bf r})}
\frac{\partial \rho_i({\bf r})}{\partial R_M}&&\nonumber\\
+ \sum_{i=A,B}\int d^3 r \rho_i({\bf r})
\frac{\partial V_i^{ext}({\bf r})}{\partial R_M}\;.&&
\end{eqnarray}

By virtue of Eq.~(\ref{eq:7}) the first sum on the right-hand side of 
Eq.~(\ref{eq:8}) vanishes whereas the second term
yields $\beta^{-1} 4\pi R_M^2~\rho_B(R_M^+)$, where 
$\rho_B(R_M^+)$ is the contact density of the nonpermeable
component of a fluid mixture at the ideal semipermeable vesicle.
Thus the sum rule reads
\begin{equation}\label{eq:9}
\beta \left(\frac{\partial \Omega}{\partial R_M} \right)_{\mu,T}=
4\pi R_M^2\rho(R_M^+)\;,\\
\end{equation}
and is satisfied by all density functionals
within weighted-density approximation \cite{Henderson83,Henderson86,Samborski93}.

Taking the derivative of Eq.~(\ref{eq:5}) with respect to $R_M$
and inserting to Eq.~(\ref{eq:9}) the sum rule can be expressed as
\begin{equation}\label{eq:10}
\rho_B(R_{M}^+) =\beta \Pi+
\frac{2 \beta \gamma(R_{M})}{R_{M}}+\beta
\left(\frac{\partial\gamma(R_{M})}{\partial R_{M}}\right)_{\mu,T}\;.
\end{equation}
In the limit $R_M\to\infty$ only the first term of Eq.~(\ref{eq:10}) survives
and we recover the planar result \cite{Martina83,Zhou88b,Powles97,Lozada04}
that the contact value of the density profile of the nonpermeable species at 
the planar semipermeable membrane is equal to the osmotic pressure. Since $\Pi$
is always lower than $p^{(II)}$
we conclude that for the same fluid the contact value at the membrane is lower than
the sum of the contact values at a nonpermeable hard wall \cite{Bryk03a}.
One can also consider the case in which the two component
mixture is inside the vesicle and the permeable component is present outside the vesicle,
i.e. the ``reverse'' of the situation plotted in Fig.~\ref{fig:1}.
In this case the contact value of the nonpermeable component will be higher than that at the planar
membrane.

\section{Theory for Depletion Potentials}
\label{sec:3}
We consider a single big colloidal hard-sphere of diameter $\sigma_C$ immersed
in a two-component mixture of smaller hard-sphere of diameters
$\sigma_A$ and $\sigma_B$, respectively. The system is in contact with
an ideal semipermeable membrane which is permeable to the $A$ component but
nonpermeable to the $B$ component, as well as to the big sphere $C$.
We also assume that $\sigma_A<\sigma_B<\sigma_C$.
The goal is to evaluate the effective depletion potential $W$ between the big
colloidal hard-sphere and the semipermeable membrane.
In our study we use a versatile approach proposed in Refs. \cite{Goetzelmann99},\cite{Roth00}.
According to this theory the depletion potential is evaluated 
on the basis of the density profiles of the small spheres undisturbed
by the presence of the big sphere.
In the dilute limit of the big spheres ($\rho_C\to 0$ or, equivalently,
$\mu_C\to -\infty$) the depletion potential
$W$ at a point ${\bf r}$ can be expressed in
terms of the difference between the one-particle direct correlation
function {$c_{C}^{(1)}$ in the bulk (${\bf r}\to\infty$) and at ${\bf r}$
\begin{equation}\label{eq:11}
\beta W({\bf r})=\lim_{\rho_{C}\to 0}
\left(c_{C}^{(1)}({\bf r}\to\infty)- c_{C}^{(1)}({\bf r})\right)\;.
\end{equation}

A convenient route for the evaluation of $c^{(1)}_C({\bf r})$ is provided by 
density functional theory \cite{Evans79}. The focal point in this approach is the
excess (over the ideal gas) free energy, 
$F_{ex}=F-\sum_i\int d^3{\bf r}\{\rho_i({\bf r})(\ln(\rho_i({\bf r}))-1)\}$.
The one-particle direct correlation function is then directly accessible via
\begin{equation}\label{eq:12}
c_C^{(1)}({\bf r})=-\beta\frac{\delta F_{ex}}{\delta\rho_C({\bf r})}\;.
\end{equation}

For mixtures of hard-spheres the exact excess free energy formula is not known
and some approximations are required. In this work we consider two approximate
functionals. The first one is the low density limit functional
\begin{equation}\label{eq:13}
\beta F_{ex}=-\frac{1}{2}\sum_{i,j}\int d^3r\int d^3r' \rho_i({\bf r})
\rho_j({\bf r}')f_{ij}({\bf r}-{\bf r}')\;,
\end{equation}
where $f_{ij}$ is the Mayer bond between a particle of species $i$ and a particle of
species $j$.
In the low density limit the density profiles $\rho_A({\bf r})$ and $\rho_B({\bf r})$
tend to $\rho_A^{bulk}$ and $\rho_B^{bulk}$, respectively. 
Then from Eqs.~(\ref{eq:11})-(\ref{eq:13}) we obtain the Asakura-Oosawa approximation
for the depletion potential
\begin{equation}\label{eq:14}
\beta W^{AO}({\bf r})=-\sum_{i=A,B}\int d^3r'(\rho_i^{bulk}-\rho_i^{(II)})
f_{iC}({\bf r}-{\bf r}')\;,
\end{equation}
where $f_{iC}({\bf r}-{\bf r}')=-\Theta((\sigma_i+\sigma_C)/2-|{\bf r}-{\bf r}'|)$.
Equation (\ref{eq:14}) implies that $W^{AO}$ is identically zero for distances larger than
$\sigma_B$ from the membrane.

The second approximate functional considered in this work is 
Rosenfeld's Fundamental Measure Theory \cite{Rosenfeld89} (FMT). 
Within this approach the excess
free energy of an $N$ component hard-sphere mixture is given by
\begin{equation}\label{eq:15}
\beta  F_{ex}=\int d^{3}r \;\Phi(\{n_{\alpha}\}) \;,
\end{equation}
where $n_{\alpha}$ denote weighted densities 
\begin{equation}\label{eq:16}
n_{\alpha}({\bf r})=\sum_{j=1}^{N}\int d^{3}r' \rho_j({\bf r}')~
w_{\alpha}^{(j)}({\bf r}-{\bf r}') \;.
\end{equation}
The weight functions $w^{(j)}_{\alpha}$ depend of geometrical properties of
individual species, see earlier papers for explicit formulas \cite{Rosenfeld89,Rosenfeld93}.

From a number of prescriptions for the excess free energy density 
$\Phi$, for the given problem we have chosen the elegant and inspiring White-Bear (WB)
formula \cite{Roth02,Yu02}
\begin{eqnarray}\label{eq:17}
\Phi(\{n_{\alpha}\})=-n_0 \ln (1-n_{3})+
\frac{n_{1}n_{2}-{\bm n}_{V1}\cdot
{\bm n}_{V2}}{1-n_{3}}&&\nonumber\\
+(n_2^3-3n_2{\bm n}_{V2}\cdot{\bm n}_{V2})\frac{n_{3}+
(1-n_3)^2\ln (1-n_3)}
{36\pi (n_3)^2(1-n_{3})^{2}}&&\,.
\end{eqnarray}

\begin{figure}[t]
\includegraphics[clip,width=8cm]{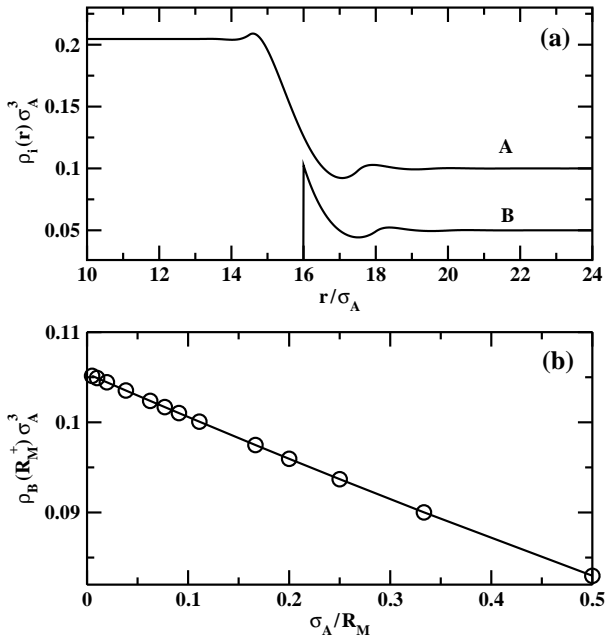}
\caption{\label{fig:2}
(a) Density profiles of a hard-sphere mixture near a semipermeable vesicle.
The bulk densities are $\rho_A^{(II)}\sigma_A^3=0.1$ and $\rho_B^{(II)}\sigma_A^3=0.05$.
(b) The contact value of the density profile of the nonpermeable component
(solid line) and the right-hand side of the sum rule, Eq.\ref{eq:10} (circles)
plotted as a function of the inverse membrane radius, $R_M$.
}
\end{figure}

As explained in \cite{Goetzelmann99,Roth00} FMT provides a very nice method of calculating 
the depletion potentials. 
In the present case of a ternary hard-sphere mixture the dilute limit
of species $C$ can be obtained by considering the set of restricted weighted densities
\begin{equation}\label{eq:18}
n_{\alpha}^{dil}({\bf r})=\sum_{j=A,B}\int d^{3}r' \rho_j({\bf r}')~
w_{\alpha}^{(j)}({\bf r}-{\bf r}') \;.
\end{equation}

The depletion potential is then obtained from
\begin{equation}\label{eq:19}
\beta W^{WB}({\bf r})=\sum_{\alpha} \int d^3 r' \Psi_{\alpha}({\bf r}';\{n_{\alpha}^{dil}\})
w_{\alpha}^{(C)}({\bf r}-{\bf r}') \;,
\end{equation}
where
\begin{equation}\label{eq:20}
\Psi_{\alpha}({\bf r}';\{n_{\alpha}^{dil}\})=\beta\left( \frac{\partial
\Phi(\{n_{\alpha}^{dil}\})}{\partial n_{\alpha}^{dil}}
\right)_{{\bf r}'}-
\beta\left( \frac{\partial
\Phi(\{n_{\alpha}^{dil}\})}{\partial n_{\alpha}^{dil}}
\right)_{\infty}\;.
\end{equation}
The computational strategy consists of two stages. 
First the profiles for the binary $A$+$B$ mixture are evaluated by minimizing the 
White-Bear version of the fundamental measure functional. The resulting 
set of restricted weighted densities is then used as an input to
a separate program, in which the depletion potentials are computed by applying
Eqs.~(\ref{eq:19})-(\ref{eq:20}). The strategy outlined above offers an important
advantage, because the computational problem is effectively reduced from three-component
to two-component mixture. This method is especially appreciated when the density profile
is a function of more than one scalar \cite{Bryk03b}.

\section{Results and Discussion} \label{sec:DFT}
\label{sec:4}
\subsection{Depletion Potentials}
Typical density profiles resulting from the WB theory are displayed in 
the upper panel of Fig.~\ref{fig:2}.
The profiles were evaluated for a binary hard sphere mixture of
sizes $\sigma_B=2\sigma_A$
near a semipermeable vesicle of the size $R_M=16\sigma_A$. The bulk densities
are $\rho_A^{(II)}\sigma_A^3=0.1$ and $\rho_B^{(II)}\sigma_A^3=0.05$.
The osmotic pressure $\beta\Pi\sigma_A^3=$0.105 391 59. The contact value of
the density profile $\rho_B(R_M^+)=$0.102 373 22 whereas the right hand-side
of Eq.~(\ref{eq:10}) is 0.102 373 78.
Similar level of agreement was found for other radii and this confirms high
accuracy of the numerical results. The plot of the contact value
as a function of the inverse membrane radius evaluated for these bulk densities
is shown in the lower panel of Fig.~\ref{fig:2}.
In the planar membrane limit, $\sigma_A/R_M=0$ the contact value tends to 
the osmotic pressure.
\begin{figure}[t]
\includegraphics[clip,width=8cm]{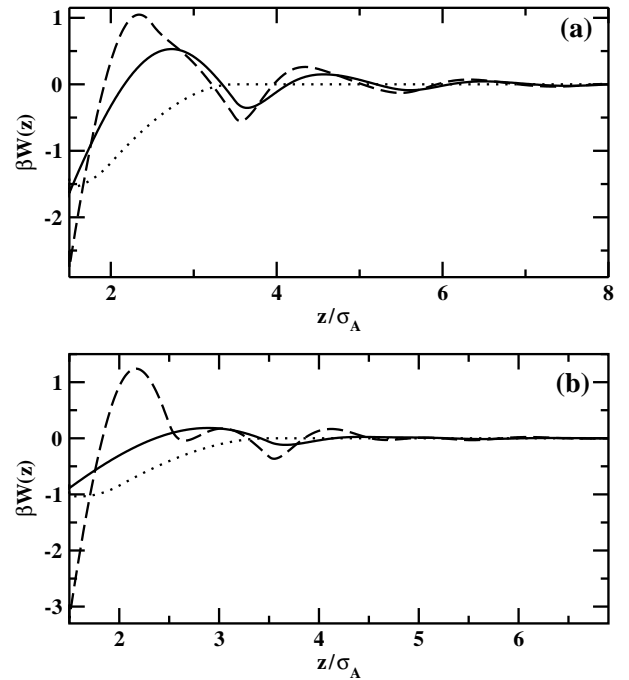}
\caption{\label{fig:3}
Depletion potentials for a single big hard sphere $C$ immersed in a
mixture of hard spheres $A$+$B$ of diameters $\sigma_B=2\sigma_A$.
The diameter of the big hard sphere
$\sigma_C=3\sigma_A$.
The solid lines denote depletion potentials between a big hard sphere and
a semipermeable planar membrane evaluated from the WB theory. The dashed lines denote
depletion potentials for the same mixture but at a hard nonpenetrable wall
evaluated from the WB theory. The dotted lines denote depletion potentials at a
semipermeable planar membrane evaluated using AO approximation.
Packing fractions are $\eta_A=0.05$, $\eta_B=0.35$ in (a) and
$\eta_A=0.15$, $\eta_B=0.25$ in (b).
}
\end{figure}

We turn now to the depletion potentials. In order to restrict a rather large
parameter space we consider the case of the fixed diameters of smaller
spheres $\sigma_B=2\sigma_A$.
Moreover, we consider the case of a constant total packing fraction
$\eta=\sum_{i=A,B}\eta_i=\sum_{i=A,B}(\pi/6)\rho_i^{(II)}\sigma_i^3=0.4$
and the case of a fixed $\eta_B=0.15$.
By varying packing fraction of the components we have a possibility to
investigate fairly dense systems but with different osmotic pressures.
Thus we hope to draw rather general conclusions. 

Figures~\ref{fig:3} and \ref{fig:4} show the depletion potentials between
a big colloidal particle of the size $\sigma_C=3\sigma_A$ and a planar
semipermeable membrane, plotted as a function
of the distance to the membrane surface $z$. The potentials were evaluated 
for the constant total packing fraction $\eta=0.4$.

\begin{figure}[t]
\includegraphics[clip,width=8cm]{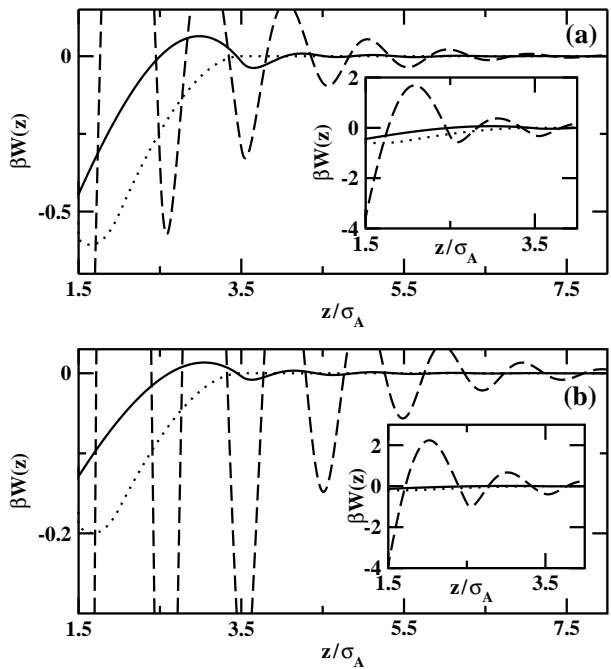}
\caption{\label{fig:4}
The same as in Fig.~\ref{fig:3} but now for packing fractions $\eta_A=0.25$, $\eta_B=0.15$ in (a)
and $\eta_A=0.35$, $\eta_B=0.05$ in (b).
The insets show the zoom-out of the main figures.
}
\end{figure}

The solid lines are the results of the WB theory whereas the dotted lines denote the results of 
the AO approximation. The dashed lines denote $W^{WB}$ but for the planar, nonpenetrable wall.
We observe that the differences between the colloid-wall and the colloid-membrane potentials are
rather significant. In all cases the effect of the membrane permeability is to decrease the effective
interactions, relative to the hard wall. The oscillations are less pronounced and the contact values
of the depletions potentials are smaller in magnitude. These trends are already well visible for small
packing fractions of the smaller spheres (cf. Fig.~\ref{fig:3}a, $\eta_A=0.05$, $\eta_B=0.35$)
and increase with $\eta_A$ (cf. Fig.~\ref{fig:3}b, $\eta_A=0.15$, $\eta_B=0.25$). For high
packing fractions of the smaller spheres and low $\eta_B$, (see Fig.~\ref{fig:4}) the difference between 
the corresponding potentials is rather dramatic. The contact value of $\beta W$ for the membrane
is less than -0.5 whereas for the hard wall this value is around -4 (see the insets to Fig.~\ref{fig:4}).
Note also that the oscillations, and in particular the first maximum, are significantly smaller for
the membrane than for the wall. 

That the depletion potential near a semipermeable membrane is weaker than the potential
near a nonpermeable wall can be anticipated already from Eq.~(\ref{eq:10}). For the one-component
hard sphere system the contact value of the depletion potential, as predicted by 
the accurate parameterization \cite{Roth00}, is linear in the packing fraction of the small spheres.
The packing fraction of the small spheres itself is related to the contact value of the profile near
the wall, thus the higher the pressure and the contact value of the density profile, 
the stronger the depletion potential. While for mixtures such
parameterization is unknown it is reasonable to assume a similar behavior. Since the contact value
near the membrane is always lower, the depletion potential is expected to be weaker and this
is confirmed by the explicit calculations.

\begin{figure}[t]
\includegraphics[clip,width=8cm]{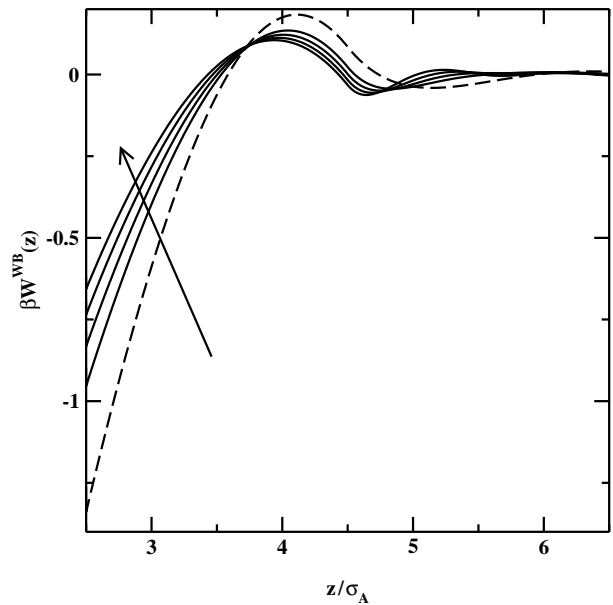}
\caption{\label{fig:5}
Depletion potentials for a single big colloidal sphere of diameter
$\sigma_C=5\sigma_A$ near a semipermeable planar membrane. The packing fraction of the
nonpermeable component $\eta_B=0.15$, and $\sigma_B=2\sigma_A$. The solid lines denote the potentials
for the packing fractions of the permeable component $\eta_A=0.10$, 0.15, 0.20, and 0.25. The arrow indicates
the increase of $\eta_A$. The dashed line indicates the result for $\eta_A=0$.
}
\end{figure}

It is interesting to note that the simple Asakura-Oosawa approximation (dotted lines if Figs.~\ref{fig:3} and
\ref{fig:4}) performs relatively well when compared with the WB theory. It is clear that the AO approximation
cannot account for the oscillatory part of the depletion potential but the contact value resulting from this 
approach does not differ too much from the accurate WB theory. 

Figure \ref{fig:5} shows the depletion potentials between a single big colloidal particle
of the size $\sigma_C=5\sigma_A$ and a semipermeable planar membrane, evaluated for a constant
packing fraction of the nonpermeable component $\eta_B=0.15$ and for different packing fractions of
the permeable component $\eta_A=0.10$, 0.15, 0.20, and 0.25. The direction of the increase of $\eta_A$ is
indicated by the arrow and it is also the direction of decreasing osmotic pressure. We observe that for constant
$\eta_B$ the effective interactions get stronger with an osmotic pressure increase (the same trend is observed in
Figs.~\ref{fig:3}-\ref{fig:4}). For reference we also show the potential for $\eta_A=0$ (dashed line) and
it is clearly the strongest potential for this $\eta_B$.
\begin{figure}[t]
\includegraphics[clip,width=8cm]{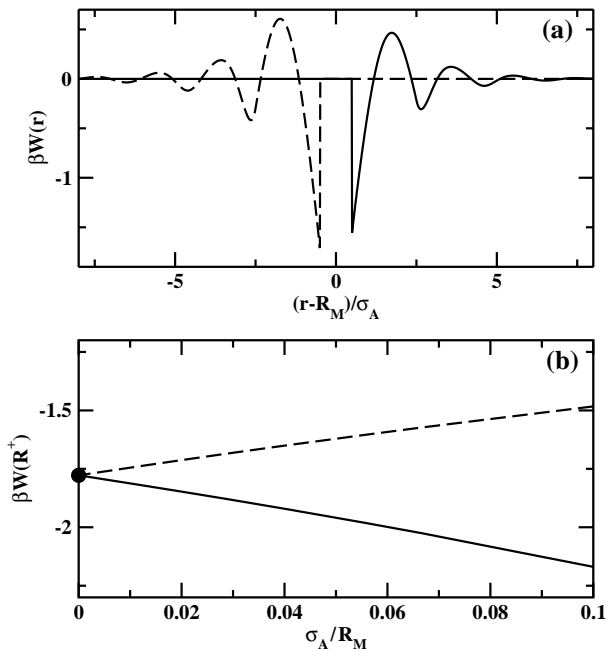}
\caption{\label{fig:6}
(a) Depletion potentials between a single big colloidal sphere of diameter
$\sigma_C=3\sigma_A$ and a semipermeable vesicle
of the radius $R_M=20\sigma_A$ plotted as a function of the relative distance to the vesicle
$(r-R_M)/\sigma_A$. 
The solid line denotes the potential for the system in which the two component part of the system is
located outside of the vesicle whereas the dashed line denotes the potential the system in which the two component
part of the system is located inside the vesicle. 
The packing fractions $\eta_A=0.05$ and $\eta_B=0.35$.
(b) Contact value of the depletion potential between a single big colloidal sphere of diameter
$\sigma_C=5\sigma_A$ and a semipermeable vesicle plotted as a function of the vesicle curvature.
The solid line denotes the potential for the system in which the two component part of the system is
located outside of the vesicle whereas the dashed line denotes the potential the system in which the two component
part of the system is located inside the vesicle. The packing fractions $\eta_A=0.30$ and $\eta_B=0.10$.
Black dot indicates the planar result limit.
}
\end{figure}

Finally in Fig.~\ref{fig:6} we show the influence of the curvature of the membrane on
the depletion potentials. We consider the depletion potential near a convex and 
a concave semipermeable membrane.
This corresponds to the cases where the two component part of the system is located outside of, and inside
the vesicle, respectively.  
Similar to the hard wall case \cite{Roth99} we find that the depletion potential inside the vesicle
is stronger, whereas the potential outside of the vesicle weaker,
than the corresponding potential at the planar membrane (cf. Fig.~\ref{fig:6}a).
However, this effect is moderated in comparison to the hard wall case because,
in general, the depletion potentials near semipermeable membranes are weaker.
Figure Fig.~\ref{fig:6}b shows the contact value of the depletion potential
at a semipermeable membrane plotted as a function of the inverse vesicle radius.
We find that, although the convergence to the planar membrane
result (marked by the black dot) is rather slow, one has to go to rather small radii in order
to noticeably increase/decrease the contact value of the depletion potential. 

\subsection{Asymptotic behavior}
The behavior of the depletion potential at large separations
can be ascertained by using the general theory of asymptotics
of pair correlations functions \cite{Evans93,Evans94}.
It was demonstrated that the decay of many structural
properties of inhomogeneous fluids such as density 
profile, solvation force and depletion potential\cite{Roth00} should be the same
as the asymptotics of the bulk pair correlation functions $h_{ij}$.
Recently Grodon {\it et al.} \cite{Grodon04,Grodon05}
presented a detailed investigation of the asymptotic decay of pair correlation
functions in binary hard-sphere mixtures. The main finding 
is that for sufficiently asymmetric mixtures there is a structural
crossover line in the ($\eta_A$,$\eta_B$) plane which separates
two distinct regimes of the asymptotic decay. Below we recall main
predictions of the general theory of asymptotics of pair correlations functions
for mixtures.

\begin{figure}[t]
\includegraphics[clip,width=8cm]{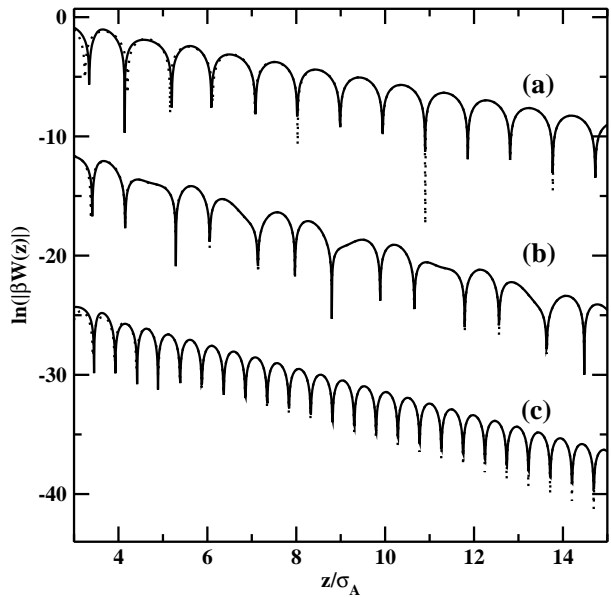}
\caption{\label{fig:7}
Asymptotic decay of the depletion potential between a single big colloidal particle
and a semipermeable planar membrane. Shown are the results of the WB theory (solid lines)
and the asymptotic forms evaluated from Eqs.~(\ref{eq:23})-(\ref{eq:24}) (dashed lines). 
$\sigma_C=3\sigma_A$ and the packing fractions are
$\eta_A=0.05$ and $\eta_B=0.35$ for the potential marked as (a), 
$\eta_A=0.1435$ and $\eta_B=0.2565$ for the potential marked as (b),
and $\eta_A=0.35$ and $\eta_B=0.05$ for the potential marked as (c).
Potentials (b) and (c) are shifted by $\ln|\beta W(z)|=$10 and 20, respectively.
}
\end{figure}

The Fourier transforms of the total correlation functions for
a binary mixture of $A$ and $B$ species can be written as
\begin{equation}\label{eq:21}
\hat{h}_{ij}(k)=\frac{\hat{N}_{ij}(k)}{\hat{D}(k)}\;\;i,j=A,B\;.
\end{equation}
While the nominators $\hat{N}_{ij}(k)$ in Eq.~(\ref{eq:21}) are different
for different pairs $ij$, all three total correlation functions have the same denominator $\hat{D}(k)$
\begin{equation}\label{eq:22}
\hat{D}(k)=[1-\rho_A \hat{c}^{(2)}_{AA}(k)][1-\rho_B \hat{c}^{(2)}_{BB}(k)]
-\rho_A\rho_B\hat{c}^{(2)}_{AB}(k)^2\;.
\end{equation}
In the above $\hat{c}^{(2)}_{ij}(k)$ is the Fourier transform of the (two-particle)
direct correlation function.
The inverse Fourier transform of Eq.~(\ref{eq:22}) is formally obtained
via the residue theorem. While in general an infinite
number of residues and the corresponding poles contributes to $h_{ij}(r)$,
the asymptotic decay is governed by the leading order pole 
$p^{(LO)}=\pm a_1+ia_0$, that is by the pole with the smallest imaginary part.
Close to a crossover point two poles $p^{(1)}=a_1+ia_0$ and $p^{(2)}=\tilde{a}_1+i\tilde{a}_0$
have similar imaginary parts and both contribute to the decay.
Provided that the Fourier transforms of  $c_{ij}(r)$ are known
analytically, the poles can be conveniently determined by finding complex solutions to the equation
$\hat{D}(k)=0$. In the present case of the binary hard-sphere mixture 
the direct correlation functions $c_{ij}(r)$ and their Fourier transforms
are readily obtained from the density functional theory 
\begin{eqnarray}\label{eq:23}
c_{ij}(r)&=&-\beta\frac{\delta^2 F_{ex}[\rho_A({\bf r}),\rho_B({\bf r})]}
{\delta\rho_i({\bf r})\delta\rho_j({\bf r}')}\nonumber\\
&\; &\;\nonumber\\
&=&-\sum_{\alpha,\beta}\frac{\partial^2\Phi(\{n_{\nu}\})}
{\partial n_{\alpha}\partial n_{\beta}}w_{\alpha}^{(i)}\otimes w_{\beta}^{(j)}\;.
\end{eqnarray}
In the above $\otimes$ denotes the convolution.
Explicit expressions for $c_{ij}(r)$ can be found in Ref.\cite{Grodon04}.

The asymptotic decay for pair correlations determines the decay of other
structural and thermodynamic quantities describing inhomogeneous fluids.
Thus the asymptotics of the depletion potential between a hard-sphere colloidal
particle and a semipermeable planar membrane, for systems away from a crossover
point, can be written as
\begin{equation}\label{eq:24}
\beta W(z)\sim A_{d}\exp(-a_0 z)\cos(a_1 z-\Theta_d)\;.
\end{equation}
The amplitude $A_d$ and the phase $\Theta_d$ of the depletion potential depend
on the external potential whereas the characteristic decay length $a_0^{-1}$ 
and the oscillation wavelength $2\pi/a_1$ are
the same as in the decay of the pair correlation functions.
However in the proximity of a crossover point a two-pole approximation is
necessary
\begin{eqnarray}\label{eq:25}
\beta W(z)&\sim& A_{d}\exp(-a_0 z)\cos(a_1 z-\Theta_d)\nonumber\\
&+&\tilde{A}_{d}\exp(-\tilde{a}_0 z)\cos(\tilde{a}_1 z-\tilde{\Theta}_d)\;.
\end{eqnarray}

In Fig.~\ref{fig:7} we show the depletion potentials between a single big hard
sphere of diameter $\sigma_C=3\sigma_A$ and a semipermeable planar membrane.
All potentials were evaluated for the constant total packing fraction of the
smaller spheres, $\eta_A+\eta_B=$0.4. The potential marked
by (a) is for $\eta_A=0.05$ and $\eta_B=0.35$, whereas the potential marked by 
(c) is for $\eta_A=0.35$ and $\eta_B=0.05$. We find that for the system (a), rich
in $B$ species, the real part of the leading order pole
$a_1=3.28257\sigma_A^{-1}$. This leads to the oscillation wavelength
1.91411$\sigma_A$, which is close to $\sigma_B$. On the other hand, for the 
system (c), rich in $A$ species, $a_1=6.41187\sigma_A^{-1}$ and this leads to
the oscillation wavelength 0.97993$\sigma_A$. The asymptotic form 
[cf. Eq.~(\ref{eq:24}), dashed lines in Fig.~\ref{fig:7}] agrees very well with the DFT results
(solid lines) already for separations of the order of several diameters.
The depletion potential (b) was calculated for $\eta_A=0.1435$ and 
$\eta_B=0.2565$. These packing fractions are very close to the crossover point.
We note that the asymptotic form, Eq.~(\ref{eq:25}), gives an excellent
approximation already for $z>5\sigma_A$. This is the consequence of the fact
that the asymptotic form sets in at intermediate separations. We conclude that
the asymptotics of the depletion potential near a semipermeable membrane is the
same as for the potential of the same mixture near a nonpermeable wall.

\section{Conclusions} \label{sec:conclusion}
\label{sec:5}
Within the framework of density functional theory we have investigated
the effect of the membrane permeability on depletion
potentials in colloidal systems. We have focused on highly idealized systems
and considered hard sphere colloids and an infinitely thin ideal membrane.
The main conclusion is that the colloid-semipermeable
membrane effective interactions are generally weaker than those near
a hard nonpermeable wall. For hard-sphere--ideal-membrane systems
this finding can be ascertained by analyzing statistical-mechanical sum rules
which relate the osmotic pressure to the contact value of the density profile
of the nonpermeable component.
We have found that the depletion potential gets stronger with an increase
of the packing fraction of the nonpermeable species and with the osmotic pressure.
The structure of the depletion potential and its asymptotics is the same
as for the hard wall systems. In particular, we have found the structural
crossover behavior that separates two distinct asymptotic regimes of the
depletion potential. Likewise, the influence of the membrane curvature on the 
effective potential is similar to the hard-wall case but the effect is moderated.
Although no comparison with computer simulation is presented we expect our
White-Bear theory results to be very accurate.

It follows from the present study that in order to induce sufficiently strong 
colloid-membrane attraction the membrane should have high enough curvature.
This effect can be brought about by changing locally the shape of the membrane
in the near vicinity of a colloidal particle and our results underline the importance
of this process. Another possibility is to increase the osmotic pressure.

\begin{acknowledgments}
This work has been supported by KBN of Poland under the Grant
3T09A 069 27 (years 2004-2006).
\end{acknowledgments}


\end{document}